# Segmentation Based Approach to Dynamic Page Construction from Search Engine Results

K.S. Kuppusamy, G.Aghila
Department of Computer Science, School of Engineering & Technology
Pondicherry University
Pondicherry, India

***Abstract*— The results rendered by the search engines are mostly a linear snippet list. With the prolific increase in the dynamism of web pages there is a need for enhanced result lists from search engines in order to cope-up with the expectations of the users. This paper proposes a model for dynamic construction of a resultant page from various results fetched by the search engine, based on the web page segmentation approach. With the incorporation of personalization through user profile during the candidate segment selection, the enriched resultant page is constructed. The benefits of this approach include instant, one-shot navigation to relevant portions from various result items, in contrast to a linear page-by-page visit approach. The experiments conducted on the prototype model with various levels of users, quantifies the improvements in terms of amount of relevant information fetched.**

***Keywords-information retrieval; search engine results; personalization;***

## I. Introduction

The web search engines have become one of the most used tools in the World Wide Web. The mammoth size of the web and the dynamic nature of sites have fueled this growth in usage of search engines. The visits to many urls are now made from the result links rather than directly entering them in to the address bar of web browsers. Though the web search engines includes various components like crawler, indexer and searcher the result rendering module plays a vital role in the success of search engines since the result rendering is the module where the end-user interaction happens.

The growth of resources in the World Wide Web is at a very rapid pace [1]. Another factor to consider in this regard is the speed at which these resources are added, deleted and modified. The study [2] indicates that the rate at which this process happens is variable across the resources. As the state of World Wide Web is getting changed very dynamically the search engine's result cannot be an exception to this phenomenon.

This paper proposes an innovative approach in the result rendering process of search engines, by transforming them into a more dynamic entity rather than being a static snippet list. The objectives of this research work are as listed below:

- Proposing a model for dynamic page construction from the search engine results using web page segmentation based approach.
- Validation of the proposed model using prototype implementation and conducting experiments with various levels of users.

The remainder of this paper is organized as follows. In Section II motivations for this research work are listed. Section III would illustrate the model and algorithms for dynamic page construction from search engine results using segmentation approach. Section IV is about the experimental setup and the results. In Section V the conclusions and future directions are given.

## II. Motivations

This section would highlight various research works related to the theme of this paper. The idea presented in this paper involves following active research topics in the information retrieval process.

- Web Page Segmentation
- Search Engine result visualization.
- Personalization

### *A. Web Page segmentation*

Web page segmentation is an active research topic in the information retrieval domain in which a wide range of experiments are conducted. Web page segmentation is the process of dividing a web page into smaller units based on various criteria. The following are four basic types of web page segmentation method:

- Fixed length page segmentation
- DOM based page segmentation
- Vision based page segmentation
- Combined / Hybrid method

A comparative study among all these four types of segmentation is illustrated in [3]. Each of above mentioned segmentation methods have been studied in detail in the literature. Fixed length page segmentation is simple and less complex in terms of implementation but the major problem with this approach is that it doesn't consider any semantics of the page while segmenting. In DOM base page segmentation, the HTML tag tree's Document Object Model would be used while segmenting. An arbitrary passages based approach is given in [4]. Vision based page segmentation (VIPS) is in parallel lines with the way, humans views a page. VIPS [5] is a popular segmentation algorithm which segments a page based on various visual features.

Apart from the above mentioned segmentation methods a few novel approaches have been evolved during the last few years. An image processing based segmentation approach is illustrated in [6]. The segmentation process based text density of the contents is explained in [7]. The graph theory based approach to segmentation is presented in [8].

### *B. Search Engine Result Visualization*

The result visualization of search engines is yet another active research topic. Various efforts are on to make the search engine results to be displayed in a more interactive form. Some of the efforts incorporate the context sensitiveness in the search process itself [9]. This facilitates an improved searching context for the end-user. Clustering of search results into various groups is another approach in the result visualization process. [10], [11] are examples for such clustering based engines. In the clustering approach, instead of displaying the results as a linear list of snippets they would be grouped under various clusters. This facilitates the user to navigate the results in a much easier and effective manner.

The research work [12] aims at semantically selecting the snippets. In this work, instead of just picking up the sentences where the first occurrence of keywords occurs, the portion of the page to be selected as a part of the snippet text would be decided semantically. This is performed by the Semantic Snippet Selection (SemSS) algorithm. This would make the user to easily judge whether a particular result item is to be clicked or not.

### *C. Personalization*

Personalization is the process of customizing based on the user requirements and preferences. There exist many research works to personalize the search result rendering. The work presented in [13], proposes a method which utilizes the search experiences of the earlier searchers. Generally, the personalized result rendering is based upon the "feedback" from the end-users. There exist two types of feedbacks in the search engines. They are as listed below:

- Explicit Feedback
- Implicit Feedback

In the explicit feedback mechanism user has to explicitly indicate the relevant and non-relevant items. In the case of implicit feedback it would gathered automatically based on the actions performed by the user. Here the user is not required to explicitly mark it as relevant or not relevant. Both these types of feedbacks are discussed in [14], [15], [16].

The approach followed in the proposed model harness the combinatorial benefits of using web page segmentation, web search engine result visualization and personalization.

## III. THE MODEL

This section would elaborate the proposed segmentation based model for dynamic web page construction from search engine results. This model is illustrated in Fig.1.

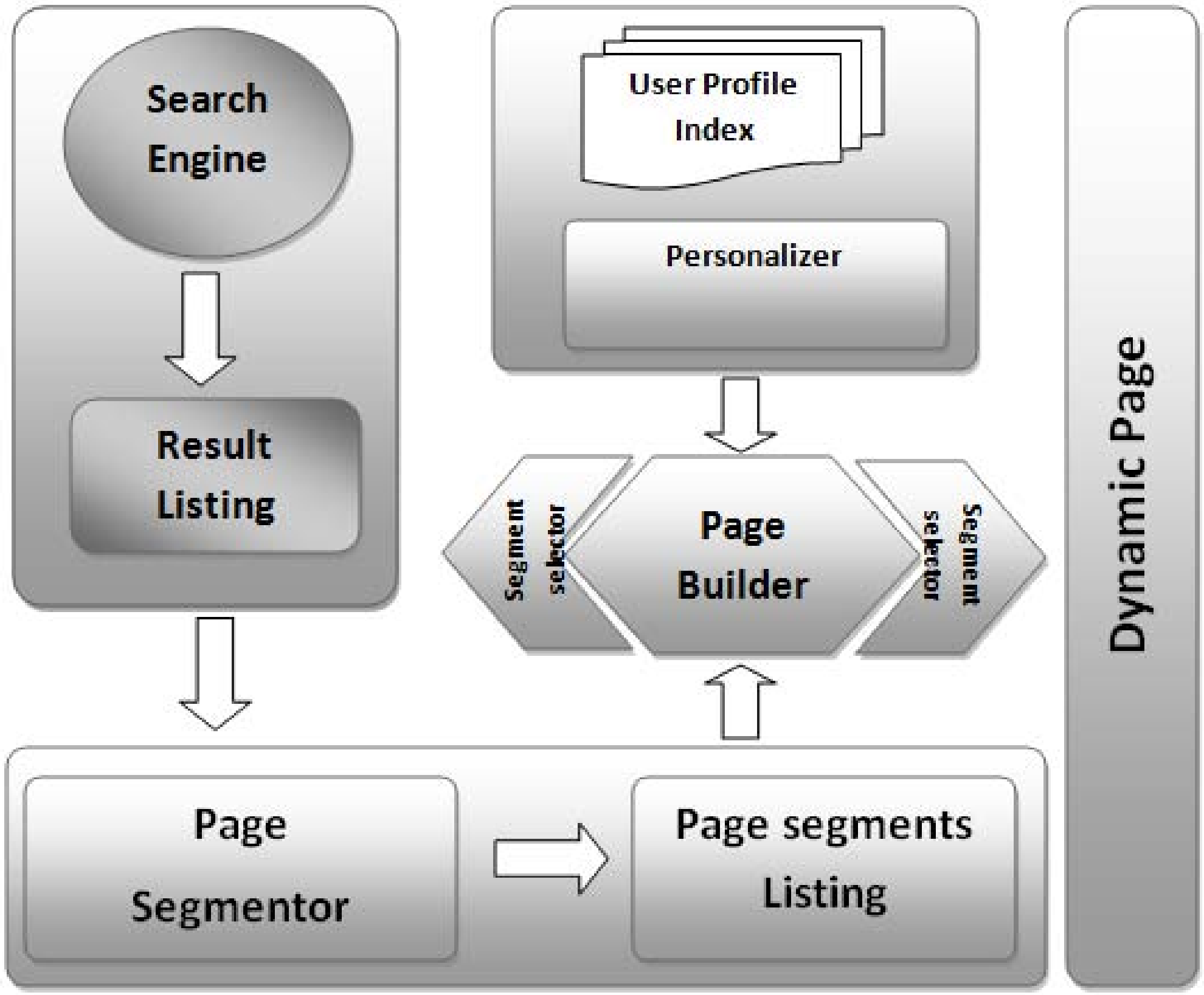

Figure 1. The model for dynamic page construction from search engine results

### A. *The mathematical Model*

The process starts with the search engine fetching results for a given query. In the proposed model this search engine component is assumed as a readily available service. This service would return the results array for any user supplied query. The second component is page segmentor. The role of page segmentor is as listed below:

- Fetch corresponding pages from first N search results.
- Apply the segmentation algorithm on each of them
- Create a segmentation matrix from the search result array.

The page segmentor would fetch the pages from the World Wide Web and apply the segmentation algorithm on it. The techniques listed in the previous section can be used for this purpose. Hence this paper approaches the resultant web page from the user prospective; the algorithm for segmentation listed in [5] which harnesses both the semantic and visual aspects of the web page can be used effectively for our purpose.

Let us represent the result list as Ψ. This result list Ψ would hold 'n' number of pages.

$$\Psi = \{\rho_1, \rho_2, \rho_{3\,\ldots}, \rho_\nu\} \quad (1)$$

Each of these result pages $\rho_1$ would be segmented in to semantically and visually related items. They are represented as shown in (2).

$$\rho = \{ \varepsilon_1, \varepsilon_2, \ldots \varepsilon_\kappa\} \quad (2)$$

Upon completion of the segmentation process the segments matrix is created. This segment matrix would be supplied as input to the page builder.

$$\Omega = \begin{bmatrix} \varepsilon_{11} & \varepsilon_{12} & \varepsilon_{1n} \\ \varepsilon_{21} & \varepsilon_{22} & \varepsilon_{2n} \\ & & \\ \varepsilon_{m1} & \varepsilon_{m2} & \varepsilon_{mn} \end{bmatrix} \quad (3)$$

The page builder module performs the following tasks.

- Evaluate each of the elements in the segments matrix
- Assign a weight score for each segment as a result of above step and create the segment-weight matrix

$$\Upsilon = \begin{bmatrix} |\varepsilon_{11}| & |\varepsilon_{12}| & |\varepsilon_{1n}| \\ |\varepsilon_{21}| & |\varepsilon_{22}| & |\varepsilon_{2n}| \\ & & \\ |\varepsilon_{m1}| & |\varepsilon_{m2}| & |\varepsilon_{mn}| \end{bmatrix} \quad (4)$$

- Enhance the segment-weight matrix with profile index terms

$$M = \{\eta_1, \eta_2 ... \eta_n\} \quad (5)$$

Where each item in M indicates the user profile keywords

$$\Phi = \frac{\begin{bmatrix} |\varepsilon_{11}| & |\varepsilon_{12}| & |\varepsilon_{1n}| \\ |\varepsilon_{21}| & |\varepsilon_{22}| & |\varepsilon_{2n}| \\ & & \\ |\varepsilon_{m1}| & |\varepsilon_{m2}| & |\varepsilon_{mn}| \end{bmatrix}}{M} \quad (6)$$

- Select the resultant segments. These segments are termed as “candidate segments”.

$$\Phi_{cs} = \begin{Bmatrix} \forall\ \varepsilon_{i,j} \in \Phi : |\varepsilon_{i,j}| & if\, |\varepsilon_{i,j}|/M > \delta \\ 0 & otherwise \end{Bmatrix} \quad (7)$$

Here $\varepsilon_{i,j}$ is selected if its weight is greater than the specified threshold value $\delta$ , otherwise it is not selected. This selection process is indicated in (7).

Upon completion of construction of candidate segment matrix $\Phi_{cs}$ , the elements in the matrix are used to build the dynamic resultant web page. The pre-built templates are used for this purpose and segments would be fit into the corresponding positions by following the dynamic token replacement policies.

$$P = \begin{Bmatrix} \forall\ \varepsilon_{i,j} \in \Phi_{cs}, t_i \in T: \text{pi} = \varepsilon_{i,j} \ / \ t_i \\ \end{Bmatrix} \quad (8)$$

Each placeholder in the segment is called “token”. The token matching module would replace these tokens with the corresponding segments which we identified with the help of (7). As a result of this token replacement process, the final resultant, dynamic web page is built.

*B. The algorithm*

The algorithms to perform the dynamic page construction from the search engine results are explained in this section. For the purpose of codifying the above model two different algorithms are used. They are as listed below:

- Algorithm SegmentSelect
- Algorithm PageBuild

*1) Algorithm SegmentSelect*

The SegmentSelect algorithm is Fig.2.

```
Algorithm SegmentSelect
Input:
        Result list Ψ from search engine
        Profile Index terms M of user
Output:
        Candidate Segment Matrix Φcs

Begin
        //fetch the result pages
    1.  For each item ρ in the result list Ψ
        1.1.  Fetch the page P from web for ρi
        1.2.  Do segmentation and build segment matrix Ω

        // Build the segment-weight matrix ϒ
    2.  For each element ε in Ω
            2.1.1. Evaluate the weight for ε
            2.1.2. Add this weighted element ε to ϒ
    3.  Fetch the keywords M from profile-index list

        // calculate the personalized score
    4.  For each element ε in ϒ
        4.1.  Enhance the score of ε, ε = ε / |M|
        4.2.  Update this score to the matrix Φ

        //Build the candidate segment matrix Φcs
    5.  For each element ε in Φ
        5.1.  If (| ε| > δ) then
            5.1.1. Add ε to candidate segment matrix Φcs

End
```

Figure 2. Algorithm SegmentSelect

As shown in Fig.2. the algorithm SegmentSelect takes the result list from search engine, profile index terms as input and returns the candidate segment matrix as output.

*2) Algorithm PageBuild*

After the completion of building up of candidate segment matrix, then the algorithm PageBuild is used to build a dynamic page from the elements in the candidate segment matrix. This algorithm is illustrated in Fig. 3.

```
Algorithm  PageBuild
Input:
        Candidate Segment Matrix Φcs
        Template T
Output:
        Constructed page P.

Begin

        1.  For each token ti in template T
            1.1. Fetch segment ε from Φcs
            1.2. Replace the token ti with the segment ε
            1.3. Remove the element ε from Φcs
            1.4. Add it to the page P

        2.  If (| Φcs | > 0) then
                  repeat step 1
            else
                  return page P
End
```

Figure 3. Algorithm PageBuild

The algorithm PageBuild takes the candidate segment matrix, template as input and returns the constructed dynamic page as output. The PageBuild algorithm scans the given template T for the presence of tokens in it. For each of the tokens in the template it replaces it with an element from $\Phi_{cs}$. This process would be continued for all tokens in the template.

In cases where the number of segments in $\Phi_{cs}$ are more than the tokens in the template, it would go for another iteration of the entire loop. In this scenario the items would be appended to the existing one. This scenario is adopted in this paper by considering the fact that the presence of vertical scroll bar is more convenient to the user rather than the presence of horizontal scroll bar in the web browser scenario.

## IV. EXPERIMENTAL SETUP AND RESULTS

This section would highlight the experimental setup used for the validation of above mentioned model and algorithms. The prototype implementation is done with the software stack including Linux, Apache, MySql and PHP. For client side scripting JavaScript is used. With respect to the hardware, a dual processor system with 3 GHz of speed, 4 GB of RAM is used. The internet connection used in the experimental setup is a 64 Mbps leased line.

A screenshot of the prototype implementation is as shown in Fig.4. The resultant dynamic page for the search query “Pondicherry” is shown in the Fig.4. The user profile-index term is set as “tourism”. So the resultant dynamic page is constructed by selecting the segments from pages related to this profile-index term. The experiments were conducted in batches. For the experimentation purpose we classified the users in to three types. They were termed as

- Level I Users
- Level II Users
- Level III users

The numbers were assigned in the increasing order of their proficiency with the computers in general and World Wide Web in particular. This rationale behind such a classification is to validate the results against a spectrum of users with varying skill levels.

TABLE I. PERFORMANCE ANALYSIS

| Session | Time required to fetch the required information (in seconds) | | | | | |
|---|---|---|---|---|---|---|
| | *Level I* | | *Level II* | | *Level III* | |
| | *FEAST* | *Model* | *FEAST* | *Model* | *FEAST* | *Model* |
| 1 | 40 | 35 | 42 | 30 | 40 | 35 |
| 2 | 65 | 45 | 47 | 35 | 35 | 30 |
| 3 | 67 | 60 | 47 | 34 | 40 | 38 |
| 4 | 58 | 56 | 58 | 30 | 39 | 30 |
| 5 | 54 | 25 | 65 | 45 | 55 | 45 |
| 6 | 120 | 70 | 75 | 35 | 60 | 45 |
| 7 | 110 | 68 | 65 | 50 | 55 | 38 |
| 8 | 125 | 96 | 75 | 60 | 75 | 45 |
| 9 | 145 | 90 | 68 | 45 | 60 | 36 |
| 10 | 135 | 96 | 70 | 46 | 80 | 55 |

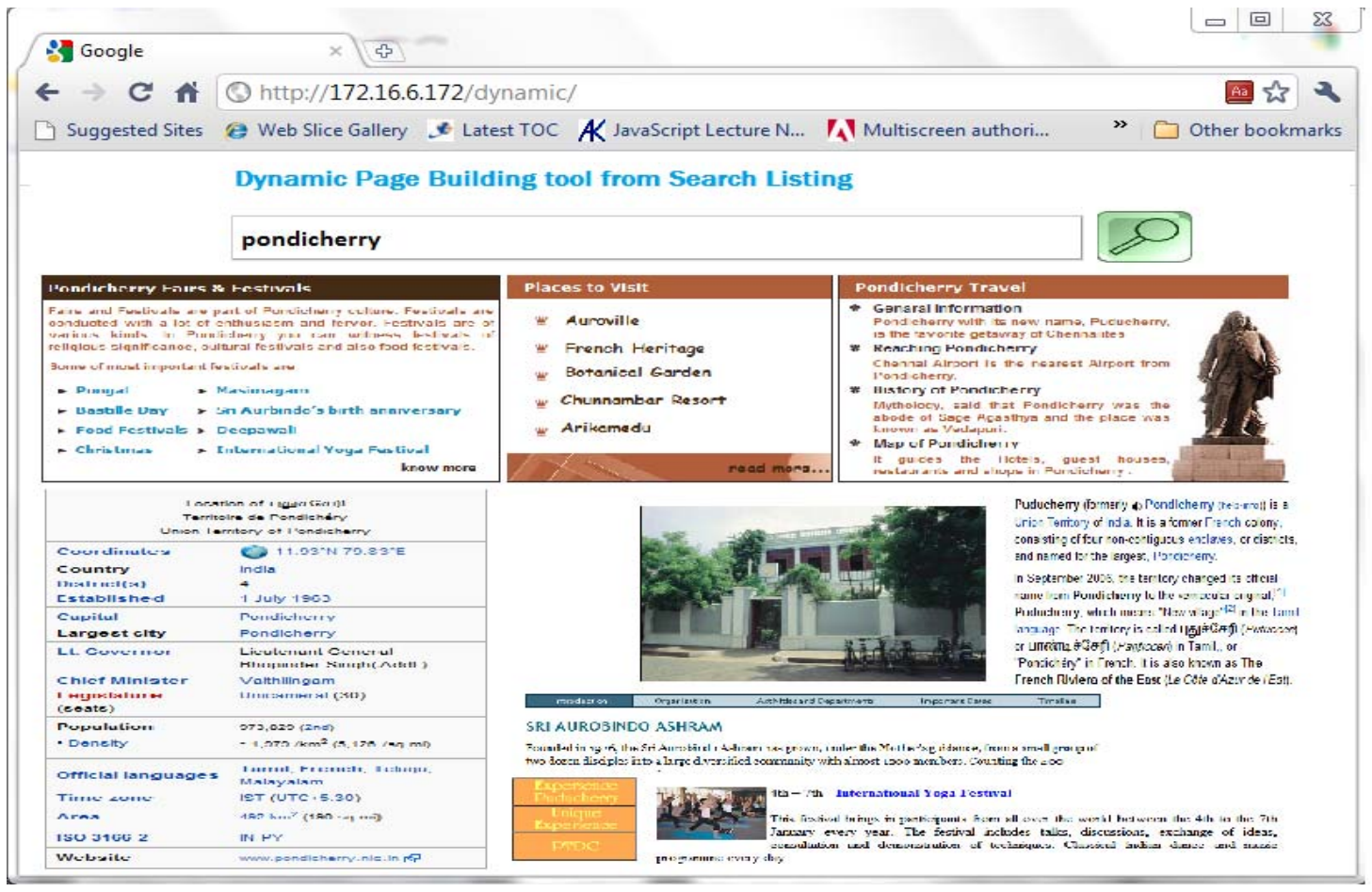

Figure 4. Screenshot of prototype implementation showing the page for the keyword “Pondicherry”

For each level of users two columns of data are provided which indicate the time required for that level user to fetch the required information using this model and FEAST (Freshness Enriched Active Searching Technology) [17]. This new model is compared with our own existing work [17] and the results are analyzed in this section.

The performance analysis of the model for Level I user is as shown in Fig.5. The t-test results are also listed in Fig.5. The values listed validate considerable improvement in the proposed model with respect to Level I users.

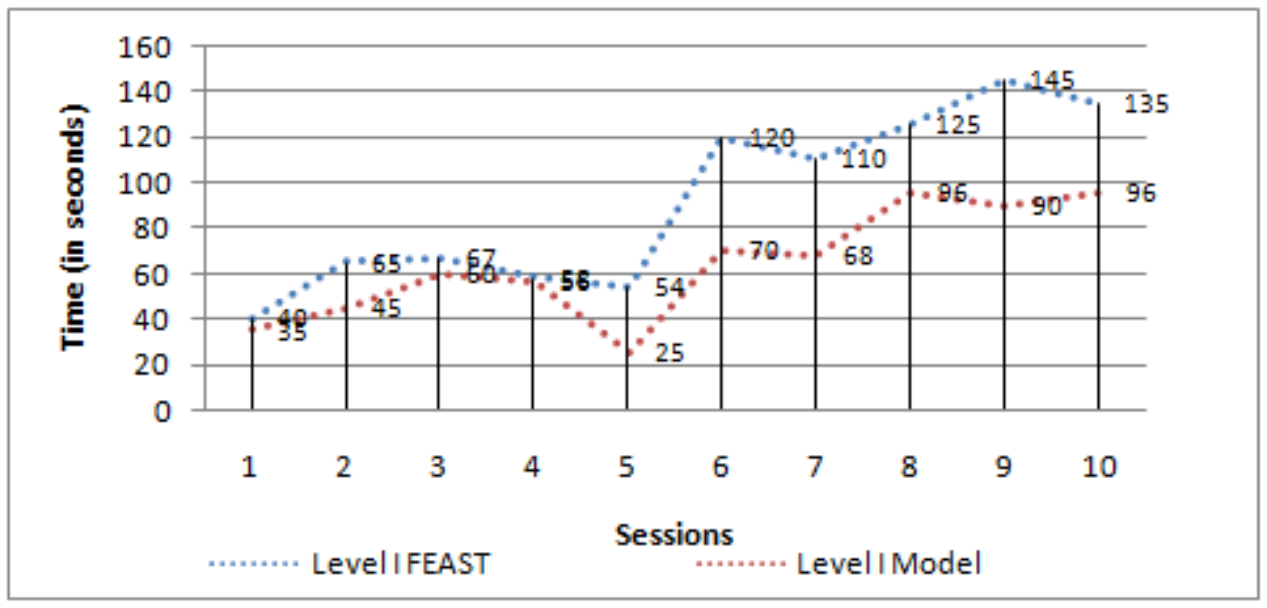

t-Test: Paired Two Sample for Means For Level I Users

| | FEAST | Model |
|---|---|---|
| Mean | 91.9 | 64.1 |
| Variance | 1501.433 | 619.8778 |
| Observations | 10 | 10 |
| Pearson Correlation | 0.912759 | |
| Hypothesized Mean Difference | 0 | |
| df | 9 | |
| t Stat | 4.63219 | |
| P(T<=t) one-tail | 0.000616 | |
| t Critical one-tail | 1.833113 | |
| P(T<=t) two-tail | 0.001233 | |
| t Critical two-tail | 2.262157 | |

Figure 5. Level I user Performance Analysis

The performance analysis of the model for Level II user is as shown in Fig.6. The t-test results are also listed in Fig.6. The values listed validate considerable improvement in the proposed model with respect to Level II users.

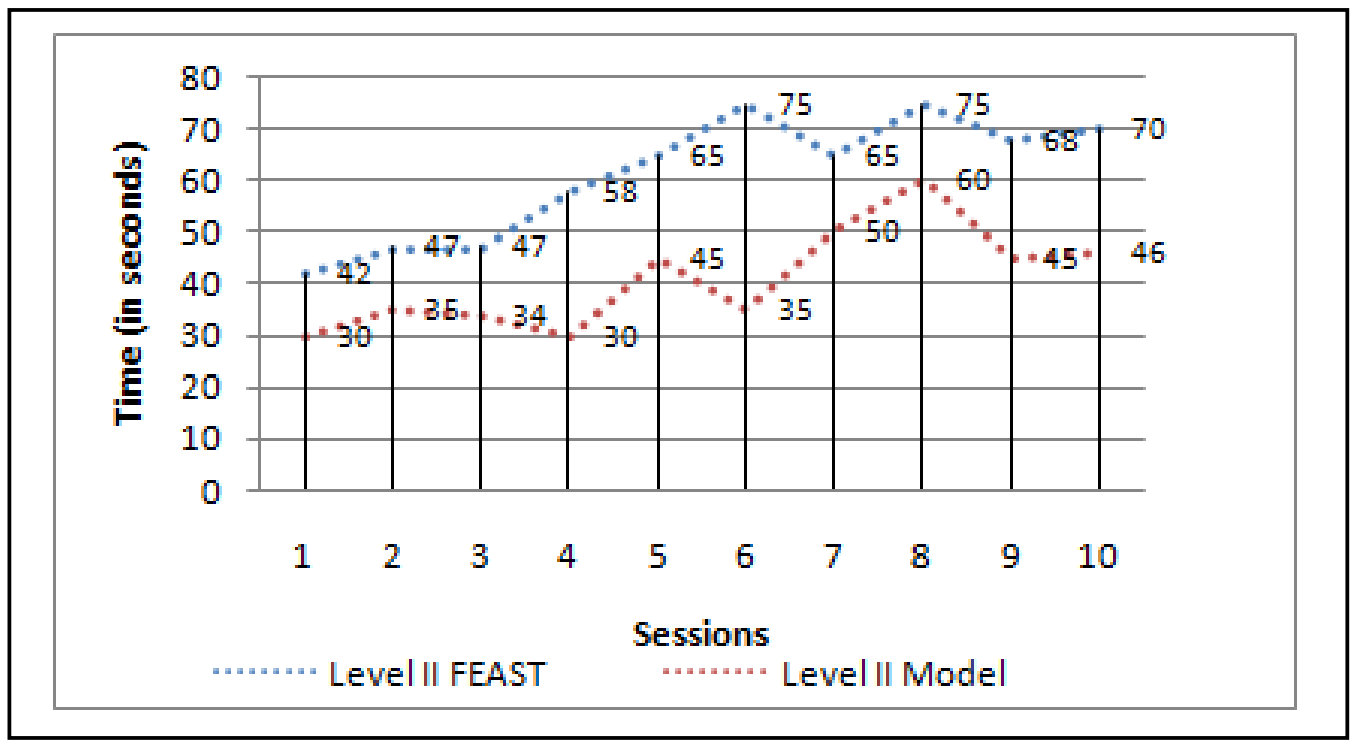

t-Test: Paired Two Sample for Means For Level II Users

| | FEAST | Model |
|---|---|---|
| Mean | 61.2 | 41 |
| Variance | 146.1778 | 95.77778 |
| Observations | 10 | 10 |
| Pearson Correlation | 0.686439 | |
| Hypothesized Mean Difference | 0 | |
| df | 9 | |
| t Stat | 7.163701 | |
| P(T<=t) one-tail | 2.64E-05 | |
| t Critical one-tail | 1.833113 | |
| P(T<=t) two-tail | 5.29E-05 | |
| t Critical two-tail | 2.262157 | |

Figure 6. Level II user Performance Analysis

The performance analysis of the model for Level III user is as shown in Fig.7. The t-test results are also listed in Fig.7. The values listed validate considerable improvement in the proposed model with respect to Level III users.

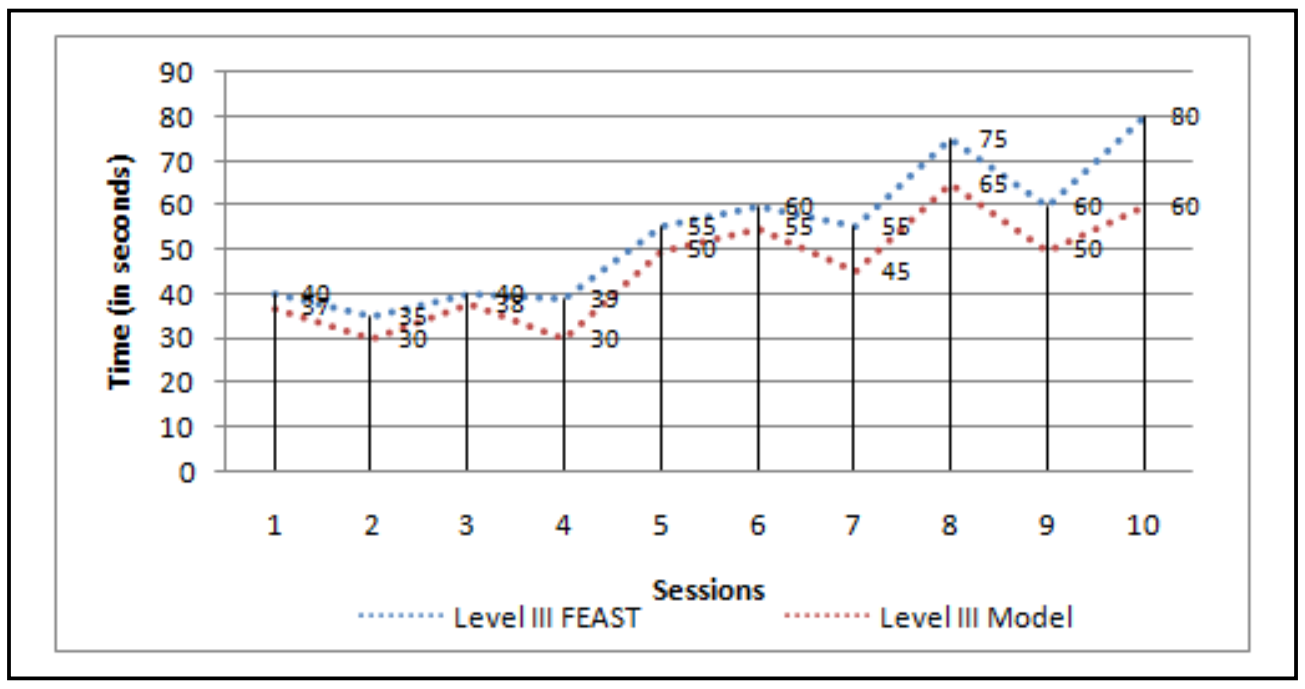

t-Test: Paired Two Sample for Means For Level III Users

| | FEAST | Model |
|---|---|---|
| Mean | 53.9 | 46 |
| Variance | 240.9889 | 147.5556 |
| Observations | 10 | 10 |
| Pearson Correlation | 0.95808 | |
| Hypothesized Mean Difference | 0 | |
| df | 9 | |
| t Stat | 4.789101 | |
| P(T<=t) one-tail | 0.000494 | |
| t Critical one-tail | 1.833113 | |
| P(T<=t) two-tail | 0.000989 | |
| t Critical two-tail | 2.262157 | |

Figure 7. Level III user Performance Analysis

The experimental results with the help of t-test clearly express the fact that the proposed model helps the user to retrieve the required information in less time.

## V. CONCLUSIONS AND FUTURE DIRECTIONS

The proposed model and its prototype implementation validates the fact that the approach followed in this paper can make the process of retrieving the required information efficiently when comparing with the linear search result list based approach. The approach caters the needs of all three types of users which have been substantiated with experimental results.

The future direction for this research includes the following:

- Improving the profile index store by incorporating ontology based techniques which would further enrich the personalization process.
- The template management can be extended so that the dynamic template selection can be achieved based on the segments retrieved.

AUTHORS PROFILE

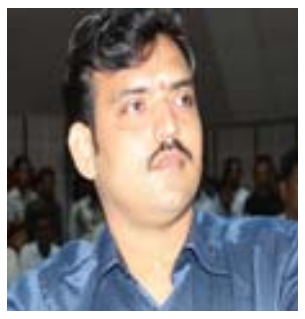

K.S.Kuppusamy is an Assistant Professor at Department of Computer Science, School of Engineering and Technology, Pondicherry University, Pondicherry, India. He has obtained his Masters degree in Computer Science and Information Technology from Madurai Kamaraj University. He is currently pursuing his Ph.D in the field of Intelligent Information Management. His research interest includes Web Search Engines, Semantic Web.

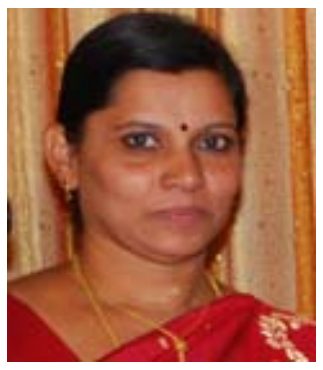

G. Aghila is a Professor at Department of Computer Science, School of Engineering and Technology, Pondicherry University, Pondicherry, India. She has got a total of 20 years of teaching experience. She has received her M.E (Computer Science and Engineering) and Ph.D. from Anna University, Chennai, India. She has published nearly 40 research papers in web crawlers, ontology based information retrieval. She is currently a supervisor guiding 8 Ph.D. scholars. She was in receipt of Schrneiger award. She is an expert in ontology development. Her area of interest include Intelligent Information Management, artificial intelligence, text mining and semantic web technologies